\documentclass[conference]{IEEEtran}
\IEEEoverridecommandlockouts 

\usepackage[cmex10]{amsmath}  
\usepackage{amssymb}
\usepackage{graphicx}   

\usepackage{ifxetex}
\ifxetex
\usepackage{xunicode}
\else
\usepackage[T1]{fontenc}
\usepackage[utf8]{inputenc}
\fi

\usepackage[noadjust]{cite}  

\newtheorem{theorem}{Theorem}

\newtheorem{lemma}{Lemma}

\newtheorem{proposition}{Proposition}

\graphicspath{{../pictures/}}

\newcommand{\figgap}{\vspace{-1.4ex}}

\newcommand{\mais}{\textsf{MAIS}}

\begin{document}

\title{A New Class of Index Coding Instances Where Linear Coding is Optimal}
\author{\IEEEauthorblockN{Lawrence Ong}
\IEEEauthorblockA{School of Electrical Engineering and Computer Science, 
The University of Newcastle, Australia
\vspace{-1.65ex}
}
\thanks{Lawrence Ong is the recipient of an Australian Research Council Discovery Early Career Researcher Award  (project number DE120100246).}
}
\maketitle

\begin{abstract}
We study index-coding problems (one sender broadcasting messages to multiple receivers) where each message is requested by one receiver, and each receiver may know some messages a priori. This type of index-coding problems can be fully described by directed graphs. The aim is to find the minimum codelength that the sender needs to transmit in order to simultaneously satisfy all receivers' requests. For any directed graph, we show that if a maximum acyclic induced subgraph (MAIS) is obtained by removing two or fewer vertices from the graph, then the minimum codelength (i.e., the solution to the index-coding problem) equals the number of vertices in the MAIS, and linear codes are optimal for this index-coding problem. Our result increases the set of index-coding problems for which linear index codes are proven to be optimal.
\end{abstract}

\section{Introduction}

We consider index-coding problems, first introduced by Birk and Kol~\cite{birkkol2006long}, where a sender communicates with multiple receivers simultaneously through a shared broadcast medium. The aim is to find the shortest codeword that the sender needs to broadcast in order for each receiver, knowing some of the  messages broadcast by the sender a priori, to obtain its requested message. Index-coding problems have been receiving much attention lately due to its equivalence to network-coding problems~\cite{rouayhebsprintsongeorghiades10,effrosrouayheb13,wonglangberg13}.




Each index-coding problem instance can be fully described by a directed or an undirected graph. 
Bar-Yossef et al.~\cite{baryossefbirk11} characterized the optimal index codelength for graphs of the following types: (a) directed and acyclic, (b) undirected and perfect, (c) undirected odd holes of five or more vertices, and (d) undirected odd anti-holes of five or more vertices. In general, the index-coding problem remains open to date, though lower and upper bounds have been obtained~\cite{baryossefbirk11,neelytehranizhang12,tehranidimakisneely12,shanmugamdimakislangberg13,yuneely13}. A lower bound is given by the number of vertices in a maximum acyclic induced subgraph (MAIS)~\cite{baryossefbirk11}. The \emph{minrank} function~\cite{baryossefbirk11} of the graph gives an upper bound (i.e., achievability), and it also gives the optimal \textit{linear} index codelength.
Both the MAIS lower bound and the minrank upper bound are NP-hard to compute~\cite{karp72,peeters96}, and both have been shown to be loose in some instances~\cite{baryossefbirk11,lubertzkystav09}. This implies that linear index codes, though having practical advantages of simplifying encoding and decoding, are not necessarily optimal.

In this paper, we extend existing results to a new class of problem instances: we show that if an MAIS is formed by removing two or fewer vertices, then the MAIS lower bound is achievable using linear index codes, meaning that linear index codes are optimal for this class of index-coding problems.   To this end, we show that this class of graphs must contain some special configurations; by proposing a new coding scheme on these special configurations, we are able to construct the required optimal index code.
This incidentally characterizes a class of (infinitely many) graphs where the minrank upper bound and the MAIS lower bound coincide.


\section{Notation and An Existing Lower Bound to The Optimal Index Codelength}

Let there be $n$ receivers, $\{1, 2, \dotsc, n\}$, and each receiver~$i$ requests a message $x_i \in \mathcal{X} \triangleq \{0,1,\dotsc, |\mathcal{X}|-1\}$ from the sender. The sender knows all the messages, $\boldsymbol{x} = (x_1,x_2,\dotsc, x_n)$. It encodes $\boldsymbol{x}$ into a length-$\ell$ codeword $\mathbb{E}(\boldsymbol{x}) \in \mathcal{X}^\ell$. It then broadcasts the codeword to all receivers noiselessly to allow each receiver~$i$---who knows some prior side information $\mathcal{K}_i \subseteq \{x_1,x_2,\dotsc, x_n\}\setminus \{x_i\}$---to decode its requested message, i.e., $\mathbb{D}_i(\mathbb{E}(\boldsymbol{x}), \mathcal{K}_i) = x_i$, for all $i \in \{1,2,\dotsc, n\}$. Here, $\mathbb{E}(\boldsymbol{x})$ is the index code, and $\ell$ is the index codelength. The aim is the find the minimum index codelength, denoted by $\ell^*$.

Each index-coding problem instance is completely specified by $\mathcal{K}_i$ for all $i \in \{1,2,\dotsc, n\}$. It can also be fully described by a directed graph $G$, consisting of a set of vertices $V(G) = \{1,2,\dotsc, n\}$ and a set of arcs $A(G)$. An arc from vertex $i$ to vertex $j$ exists, denoted by $(i \rightarrow j) \in A(G)$, if and only if receiver~$i$ knows $x_j$, or equivalently, $x_j \in \mathcal{K}_i$. By definition, there is no self loop or parallel arc. For an arc $(i \rightarrow j)$, vertex $i$ is the \emph{tail} and $j$ the \emph{head}. We term this graphical representation \emph{side-information graph}. Each vertex~$i$ in $G$ represents both message~$x_i$ and receiver~$i$.

For the special case where for each $(i \rightarrow j) \in A(G)$, there exists an arc $(j \rightarrow i) \in A(G)$, the index-coding problem instance can also be represented by an undirected graph $G''$, consisting of the same set of vertices $V(G'') = V(G)$, and a set of edges $E(G'')$, where $(i,j) \in E(G'')$ if and only if $(i \rightarrow j) \in A(G)$, which also means $(j \rightarrow i) \in A(G)$. 

We denote by $\ell^*(G)$ the optimal (or minimum) index codelength of an index-coding problem instance represented by graph $G$, which can be directed or undirected. Bar-Yossef et al.\ have shown the following lower bound~\cite{baryossefbirk11}:
\begin{equation}
\ell^*(G) \geq \mais(G),
\end{equation}
where $\mais(G)$ is the number of vertices in a maximum acyclic induced subgraph (MAIS) of $G$. An MAIS is obtained by finding an acyclic vertex-induced subgraph that has the largest number of vertices. 

For any $G$, denote an MAIS by $G'$, and the set of vertices removed by $V_\text{r} = V(G) \setminus V(G')$, where  $|V_\text{r}|$ is the number of removed vertices. Note that $|V_\text{r}|$ for each $G$ is fixed, but the MAIS may not be unique. For a \emph{vertex-induced} subgraph, when we remove a vertex $y$ from a graph $G$, we remove all outgoing arcs from and all incoming arcs to $y$, but we must keep all other remaining arcs in $G$.

\section{Main Results}

In this paper, we derive the optimal index codelength, together with the corresponding optimal index code, for a class of graphs, for any message alphabet size $|\mathcal{X}| \geq 2$. Recall that for any directed graph $G$, the number of vertices we need to remove to obtain an MAIS is denoted by $|V_\text{r}|$. With this, we now state the main result of this paper.


\begin{theorem} \label{theorem-2}
For any $G$ where $|V_\text{r}| \leq 2$, the optimal index codelength is given by
\begin{equation} 
\ell^*(G) = \mais(G) \triangleq |V(G)| - |V_\text{r}|. \label{eq:optimal-codelength}
\end{equation}
\end{theorem}


\begin{figure}[t]
\centering
\includegraphics[width=7.85cm]{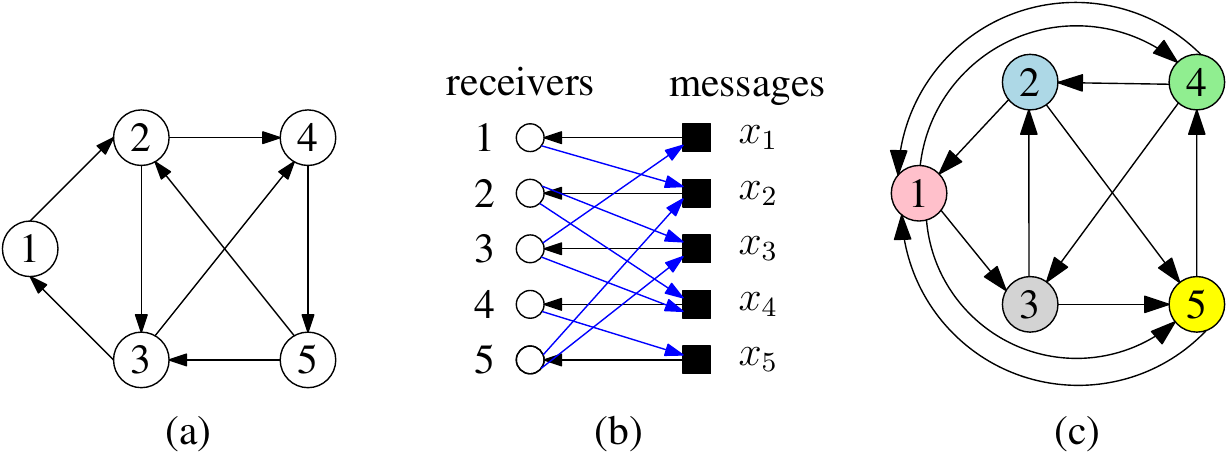}
\caption{The same index-coding instance represented by (a) a side-information graph, where the arrows represent what the receivers know; and (b) a bipartite graph, where the black arrows represent what the receivers want, and blue arrows what they know. Graph (c) is the complement of graph (a). Ignoring the arc direction, graph (c) is 5-chromatic. The local chromatic number is four (consider the number of colors in the out-neighborhood of vertex 1).}
\label{fig:bipartite}
\figgap
\end{figure}


Consider the index-coding problem instance depicted in Figure~\ref{fig:bipartite}(a) as an example. An MAIS can be formed by removing $V_\text{r}  = \{ 2,3\}$ (the choice is not unique). From Theorem~\ref{theorem-2}, we have $\ell^* = 3$.  An optimal index code is $(x_1 \oplus x_2, x_2 \oplus x_3 \oplus x_4, x_4 \oplus x_5)$ (which is also not unique), where $\oplus$ denotes addition modulo $|\mathcal{X}|$.  This problem instance is not covered by the results by Bar-Yossef et al.~\cite{baryossefbirk11}, as the side-information graph is neither acyclic nor undirected.


Neely et al.~\cite{neelytehranizhang12} and Tehrani et al.~\cite{tehranidimakisneely12} have shown the following achievability:
\begin{lemma}[\cite{neelytehranizhang12,tehranidimakisneely12}]\label{lemma}
If a directed graph $G$ contains $N$ vertex-disjoint cycles, then the index codelength of $|V(G)| - N$ is achievable.
\end{lemma}

For Figure~\ref{fig:bipartite}(a), their scheme only achieves $\ell = 4$, which is strictly suboptimal.

Recently, Shanmugam et al.~\cite{shanmugamdimakislangberg13} have shown that an upper bound of $\ell^*$ is given by the \textit{local chromatic number} of the complement graph of $G$, denoted by $\bar{G}$. The local chromatic number of $\bar{G}$ is the maximum number of colors in any \text{out-neighborhood}, minimized over all proper coloring of the undirected counterpart (by ignoring the arc direction) of $\bar{G}$. From Figure~\ref{fig:bipartite}(c), we see that this scheme achieves $\ell=4$.

Bipartite graphs are also used to represent a more general---in fact, the most general---class of index-coding problem instances where a message can be requested by more than one receiver (cf.\ side-information graphs). Neely et al.~\cite[Theorem~1]{neelytehranizhang12} found $\ell^*$ for all acyclic bipartite graphs; Yu and Neely~\cite{yuneely13} found  $\ell^*$  for all planar bipartite graphs.  The bipartite graph that represents our example is depicted in Figure~\ref{fig:bipartite}(b), which contains cycles. Also, ignoring the arc direction, we \textit{contract} the edges $\{(1,x_1), (1,x_2),(4,x_4),(4,x_5)\}$ to obtain a complete bipartite graph on three and three vertices, commonly denoted by $K_{3,3}$. Since any planar graph cannot contain a $K_{3,3}$ minor, Figure~\ref{fig:bipartite}(b) is not planar. 

Tehrani et al.~\cite{tehranidimakisneely12} have proposed a \emph{packet decomposition} scheme to obtain an upper bound on bipartite graphs. Achievability of the scheme was derived on the assumption that $\mathcal{X}$ is a large finite field. Even if the results can be shown to hold for smaller alphabets, the scheme can only achieve $\ell = 4$ for the instance in Figure~\ref{fig:bipartite}(a).

 Consequently, our results strictly extend the existing results. 
Recently, we built on the results of this paper to characterize $\ell^*$ for all $G$ up to five vertices~\cite{ong14isitsubmitted}.

The optimal \textit{linear} index codelength for any graph $G$ is given by its minrank value~\cite{baryossefbirk11}.
Characterizing graphs having a certain minrank value is hard; Dau et al.~\cite{dauskachekcheeisit12} managed to  characterize all undirected graphs whose minrank value is $|V(G)|-2$ or $|V(G)|-1$, and all directed graphs whose minrank value is 2 or $|V(G)|$. They are, however, unable to characterize directed graphs whose minrank value is $|V(G)|-1$ or $|V(G)|-2$. 
For any directed graph $G$ whose $\mais(G)$ equals $|V(G)|-1$ or $|V(G)|-2$, we show in this paper that linear index codes are optimal, meaning that $\mais(G)$ equals its minrank. So, we have incidentally characterized a subset of directed graphs whose minrank equals $|V(G)|-1$ or $|V(G)|-2$.



\section{Proof of Theorem~\ref{theorem-2}}

As $\mais(G)$ is a lower bound to $\ell^*$, we only need to prove achievability.
Recall that $|V_\text{r}|$ is the minimum number of vertices we need to remove from $G$ to make it acyclic, i.e., to obtain an MAIS. 

Firstly, suppose that $|V_\text{r}|$ = 0. Sending all messages uncoded achieves the index codelength $|V(G)|$, and we have \eqref{eq:optimal-codelength}.

Next, suppose that $|V_\text{r}|$ = 1. The directed graph $G$ must contain at least one cycle; otherwise, $|V_\text{r}| =0$. It follows from Lemma~\ref{lemma} that $|V(G)| - 1$ is achievable.

Lastly, $|V_\text{r}|=2$. There are two possibilities for $G$:
\begin{enumerate}
\item[(i)] There exist two vertex-disjoint cycles, or 
\item[(ii)] There are no two vertex-disjoint cycles.
\end{enumerate}
For case (i), it again follows from Lemma~\ref{lemma} that $|V(G)| - 2$ is achievable. The savings of two symbols (compared to sending all $|V(G)|$ symbols uncoded) can be achieved using a cyclic code on each disjoint cycle. For example, for a cycle of length $L$, say $1 \rightarrow 2 \rightarrow \dotsm \rightarrow L$, we send the following $(L-1)$ coded symbols, thereby saving one symbol: $(x_1 \oplus x_2, x_2 \oplus x_3, \dotsc, x_{L-1} \oplus x_L)$. Since each receiver $i$ knows at least one other symbol in the cyclic code, it can decode its required $x_i$.

\begin{figure}[t]
\centering
\includegraphics[width=7.5cm]{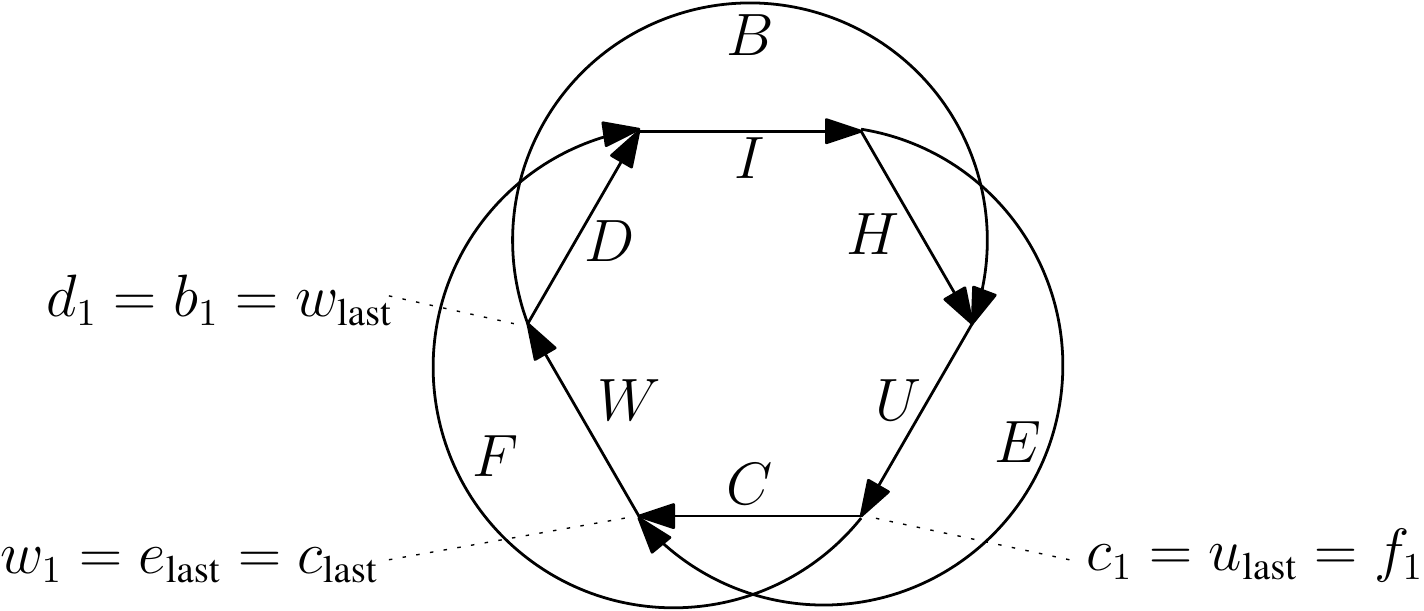}
\caption{An important element in proving Theorem~\ref{theorem-2} is to show that if $|V_\text{r}| = 2$ and condition (ii) is true, then $G$ must contain a subgraph $G_\text{sub}$ shown above. Here, every arrow represents a path, which is denoted by a capital letter. The paths do not share common vertices except the end points.
Vertices in each path is denoted by the corresponding small letter, indexed in the direction of the arcs, e.g., path $C$ is $c_1 \rightarrow c_2 \rightarrow \dotsm \rightarrow c_\text{last}$.
All paths except $I$, $W$, and $U$ must contain one or more arcs.}
\label{fig:three-cycles}
\figgap
\end{figure}

The main challenge of Theorem~\ref{theorem-2} is to show that for case (ii), even though we cannot find two vertex-disjoint cycles, we can still save two symbols. To this end, we will show that if $|V_\text{r}| = 2$ and if condition (ii) above is true, then there exists a subgraph in $G$ of a certain configuration, stated as follows:
\begin{lemma} \label{lemma:special-subgraph}
If $|V_\text{r}| = 2$, and there are no two vertex-disjoint cycles (i.e., condition (ii)), then $G$ must contain a subgraph (not necessarily an induced subgraph) shown in Figure~\ref{fig:three-cycles}.
\end{lemma}

We will design a special code on this subgraph to show that $|V(G)|-2$ is indeed achievable.
We will first present the code in the next section, and then prove Lemma~\ref{lemma:special-subgraph} in Section~\ref{sec:lemma}.


\subsection{A New Coding Scheme}

We now propose a new coding scheme that achieves the codelength $|V(G)|-2$ for case (ii). We need to show that each receiver $i \in V(G)$ can decode its intended message, i.e., $x_i$. We will propose a code for the subgraph in Figure~\ref{fig:three-cycles}, denoted by $G_\text{sub}$, and send the rest of the messages (which correspond to the vertices in $V(G) \setminus V(G_\text{sub})$) uncoded. This means all receivers $i \in V(G) \setminus V(G_\text{sub})$ can decode their intended messages, and we only need to show that all receivers $j \in V(G_\text{sub})$ can also decode their intended messages. We propose the following coding strategy: for each vertex~$a \in V(G_\text{sub})$, with all its \underline{outgoing arcs in $G_\text{sub}$} denoted by $\{(a \rightarrow a_{\text{out}1}), (a \rightarrow a_{\text{out}2}), \dotsc, (a \rightarrow a_{\text{out}T})\}$, we send the code symbol $x_a \oplus x_{a_{\text{out}1}} \oplus x_{a_{\text{out}2}} \oplus \dotsm \oplus x_{a_{\text{out}T}} \in \mathcal{X}$.

For each path in $G_\text{sub}$ (denoted by a capital letter), we denote the vertices therein by its corresponding small letter, indexed in the direction of the arcs. For example, path $C$ is  $c_1 \rightarrow c_2 \rightarrow \dotsm \rightarrow c_\text{last}$. In Figure~\ref{fig:three-cycles}, we have $c_1 = u_\text{last} = f_1$, i.e., the first vertex in path $C$ is the last vertex in path $U$, which is also the first vertex in path $F$. We use the above coding strategy to send the code symbol for all vertices in $G_\text{sub}$ except $c_1$ and $d_1$. By design, all receivers---except $c_1$ and $d_1$---can decode their requested messages, as each receiver~$a$ knows a priori the messages corresponding to the head of all outgoing arcs from $a$. So, we only need to show that receivers $c_1$ and $d_1$ can decode their respective requested messages, $x_{c_1}$ and $x_{d_1}$.

We start with receiver~$c_1 = f_1$. Knowing $x_{f_2}$ a priori, it decodes along path~$F$ (i.e., $x_{f_2} \oplus x_{f_3}, x_{f_3} \oplus x_{f_4}, \dotsc, x_{f_{\text{last}-1}} \oplus x_{f_\text{last}}$), to get $x_{f_\text{last}} = x_{i_1}$, and continues along path $I$ to get $x_{i_\text{last}} = x_{e_1}$. In the event that path $I$ has zero arc, it can also obtain $x_{e_1}$, which is $x_{f_\text{last}}$.  Also knowing $x_{c_2}$ a priori, receiver~$c_1$ decodes along path $C$ to get $x_{c_\text{last}}=x_{e_\text{last}}$,  from which it can decode backward along path $E$ to get $x_{e_2}$. Having decoded $x_{e_1}$ earlier and now $x_{e_2}$, it obtains $x_{h_2}$ from  $x_{e_1} \oplus x_{e_2} \oplus x_{h_2}$. With $x_{h_2}$, it decodes along path $H$ and then path $U$ to get $x_{u_\text{last}} = x_{c_1}$. In the event that path $U$ has zero arc, the receiver would have obtained $x_{c_1} = x_{h_\text{last}}$ earlier.

For receiver~$d_1$, knowing $x_{b_2}$, it decodes along path $B$ to get $x_{b_\text{last}} = x_{h_\text{last}}$. It then decodes backward along path $H$ to get $x_{h_2}$. Also knowing $x_{d_2}$, receiver~$d_1$ decodes along path $D$, and then path $I$ to get $x_{i_\text{last}} = x_{h_1}$. In the even that path $I$ has zero arc, receiver~$d_1$ would have obtained $x_{h_1} = x_{d_\text{last}}$ earlier. Knowing both $x_{h_1}$ and $x_{h_2}$, the receiver obtains $x_{e_2}$ from $x_{h_1} \oplus x_{h_2} \oplus x_{e_2}$. It then decodes along path $E$, and then path $W$ to get the requested $x_{d_1} = x_{w_\text{last}}$. If path $W$ has zero arc, the receiver can also obtain $x_{d_1} = x_{e_\text{last}}$.

In this coding scheme, we send one symbol for each vertex (coded symbols for $V(G_\text{sub})$, and uncoded symbols for the rest) except for $c_1$ and $d_1$. We have shown that this index code satisfies the decoding requirements of all receivers, meaning that $|V(G)|-2$ is achievable.  $\hfill \blacksquare$

\section{Proof of Lemma~\ref{lemma:special-subgraph}: A Special Configuration} \label{sec:lemma}

We first give an intuition for Lemma~\ref{lemma:special-subgraph}, by showing that there must exists three interlinked cycles in $G$, in Subsection~\ref{sec:intuition}. In Subsections~\ref{sec:figure} to \ref{sec:figure-end}, we prove that these three interlinked cycles must assume the configuration in Figure~\ref{fig:three-cycles}.

\subsection{The Existence of Three Interlinked Cycles} \label{sec:intuition}
Let $V_\text{r} = \{ u, v \}$, i.e., vertices~$u$ and $v$ are removed from $G$ to get an MAIS. We first show the following:

\begin{proposition} \label{proposition:3-cycles}
There exist three cycles in $G$, each containing either $u$, $v$, or both $u$ and $v$.
\end{proposition}

\begin{IEEEproof}
Every cycle must contain  $u$, $v$, or both. Otherwise, removing $u$ and $v$ will not give an acyclic induced subgraph.

Suppose that there is only one cycle in $G$. Removing any vertex from the cycle gives an acyclic induced subgraph. Hence, $|V(G)| - \mais(G) = 1$. (Contradiction)

Suppose that there are only two cycles in $G$. Note that these two cycles cannot be vertex-disjoint, as per condition~(ii) above. So, these two cycles must shared at least one vertex, and removing only this shared vertex gives an acyclic induced subgraph, i.e., $|V(G)| - \mais(G) = 1$. (Contradiction)

So, there must exist at least three cycles.
\end{IEEEproof}

We further show some properties of these three cycles:

\begin{proposition} \label{proposition:no-common}
There exist three cycles in $G$, where
\begin{enumerate}
\item any two cycles must have at least one common vertex, and  
\item the three cycles do not have any common vertex. 
\end{enumerate}
\end{proposition}

\begin{IEEEproof}
It follows from Proposition~\ref{proposition:3-cycles} that there are at least three cycles. As no two cycles are vertex-disjoint, we have property~1. Arbitrarily select one cycle, say $C'$. Consider every other cycle $C_k \neq C'$, and denote the set of common vertices between $C_k$ and $C'$ as $V_\text{common}(k) \triangleq V(C_k) \cap V(C')$.  Since every $C_k$ shares some vertex with $C'$, we have $V_\text{common}(k) \neq \emptyset$. 

Now suppose that $\bigcap_{\text{all } C_k \neq C'} V_\text{common}(k) \neq \emptyset$, meaning that some vertex is shared among all cycles. Then removing only this vertex from $G$ would have resulted in an acyclic subgraph (contradiction). So, there must exist two cycles, say $C_1$ and $C_2$, where $V_\text{common}(1) \cap V_\text{common}(2) = \emptyset$. Selecting $C'$, $C_1$, and $C_2$ gives property~2.
\end{IEEEproof}

Denote the subgraph formed by the three cycles in Proposition~\ref{proposition:no-common} by $G_\text{sub}$. We have the following:

\begin{proposition} \label{proposition:subgraph}
The subgraph $G_\text{sub}$, formed by the three cycles in Proposition~\ref{proposition:no-common}, satisfies both the following: (1) we cannot find two vertex-disjoint cycle in $G_\text{sub}$, and (2) we need to remove two---not fewer---vertices to make $G_\text{sub}$ acyclic.
\end{proposition}

\begin{IEEEproof}
Since $G$ cannot contain two vertex-disjoint cycles, so does any of its subgraphs. We have property~1. Denote by $N$ the minimum number of vertices we need to remove to make $G_\text{sub}$ acyclic.
From Proposition~\ref{proposition:no-common}, there is no common vertex among the three cycles. So, removing any one vertex will not disconnect all three cycles simultaneously, i.e., $N \geq 2$. On the other hand, we only need to remove two vertices, $V_\text{r}$, to make $G$ acyclic. So, removing $V_\text{r} \cap V(G_\text{sub})$ from $G_\text{sub}$ will definitely make it acyclic, i.e., $N \leq 2$. So, we have property~2.
\end{IEEEproof}

Note that these three cycles,  $G_\text{sub}$, capture all the constraints we impose on $G$ in Lemma~\ref{lemma:special-subgraph}. 

\subsection{The Three Interlinked Cycles Must Assume Figure~\ref{fig:three-cycles}} \label{sec:figure}
We will proceed to show that $G_\text{sub}$ must assume the configuration in Figure~\ref{fig:three-cycles}. 
We will build the configuration from a cycle, say $C_1$, in $G_\text{sub}$. We call it the \emph{center cycle}. We re-label the vertices in $G_\text{sub}$ such that the vertices in $C_1$ are in ascending order in the direction of the arcs, i.e., $1 \rightarrow 2 \rightarrow \dotsm \rightarrow (|V(C_1)|-1)  \rightarrow |V(C_1)| \rightarrow 1$, where the choice of vertex~1 is arbitrary. 

For any path $P$ that originates from vertex $b$ and terminates at vertex $c$, i.e., $b \rightarrow \dotsm \rightarrow c$, we refer to all $\{z: z \in V(P) \setminus \{b, c\}\}$ as \emph{inner vertices}. Here, we allow $b = c$; in such a case, $P$ is a cycle.

 We first show the following:
\begin{proposition} \label{proposition:outside-cycle}
Consider the subgraph $G_\text{sub}$ and the cycle $C_1$ in the subgraph. Every arc not in $C_1$ belongs to some \emph{outer path}, defined as a path that originates from a vertex in $C_1$ and terminates at a vertex (which can be the same vertex) in $C_1$, but with all arcs and all inner vertices (if exists) not in $C_1$.
\end{proposition}

\begin{IEEEproof}
Since $G_\text{sub}$ is constructed by three cycles, any arc, say $(i \rightarrow j)$, not in $C_1$ must belong to either $C_2$ or $C_3$ (or both). Furthermore, from Proposition~\ref{proposition:no-common}, $C_2$ and $C_3$ must each share some vertex with $C_1$. Hence, $(i \rightarrow j)$ must belong to an outer path that originates from $C_1$ and terminates at $C_1$. 
\end{IEEEproof}

Note that the outer paths cannot form any cycle outside $C_1$. Otherwise, we have two vertex-disjoint cycles. 

\begin{figure}[t]
\centering
\includegraphics[width=\linewidth]{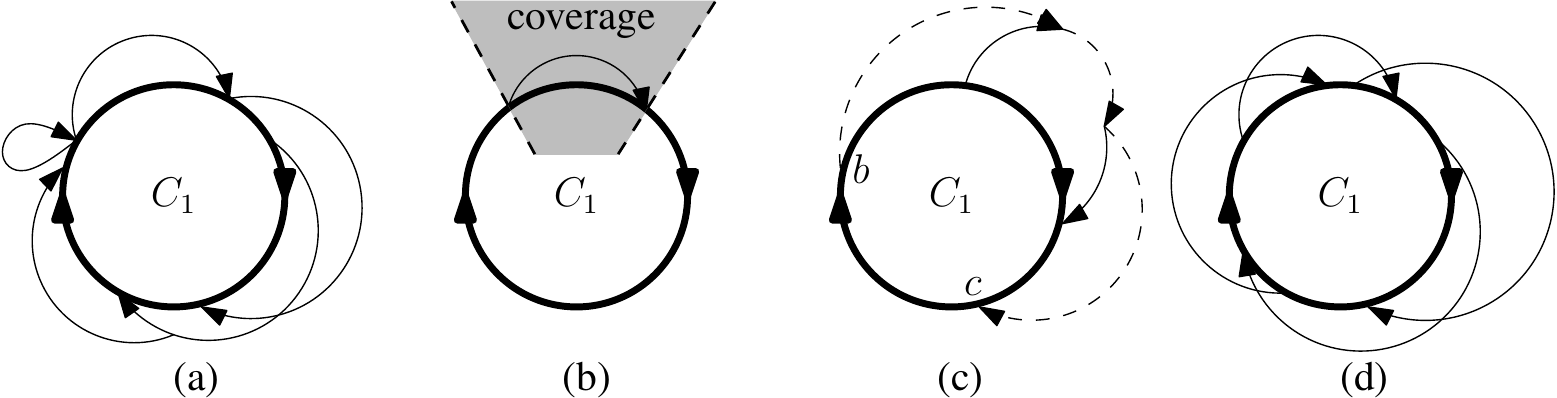}
\caption{We can always draw $G_\text{sub}$ as in subfigure~(a), i.e., a center cycle $C_1$ and outer paths from $C_1$ and back to $C_1$. Subfigure~(b) shows the coverage of an outer path, i.e., vertices in $C_1$ in the gray area \emph{excluding} the two end points. Subfigure~(c) shows that when multiple outer paths originate from one vertex, we consider only the outer path with the largest coverage, i.e., the dotted path from $b$ to $c$. The outer paths in subfigure~(d) provide full coverage.}
\label{fig:redraw}
\figgap
\end{figure}

It follows from Proposition~\ref{proposition:outside-cycle} that $G_\text{sub}$ consists of only a cycle $C_1$ and outer paths (from $C_1$ and back to $C_1$).
 Figure~\ref{fig:redraw}(a) shows an example of $G_\text{sub}$ where $C_1$ is marked with thick arrows and all outer paths thin arrows. 

We now prove a key proposition for proving Lemma~\ref{lemma:special-subgraph}.
\begin{proposition} \label{proposition:outer-path}
Remove vertex~1 in $C_1$. There exists another cycle in $G_\text{sub}$ if and only if there is an outer path from some $b \in V(C_1) \setminus \{1\}$ to some $c \in V(C_1) \setminus \{1\}$, where $b \geq c$.
\end{proposition}

\begin{figure}[t]
\centering
\includegraphics[width=8.3cm]{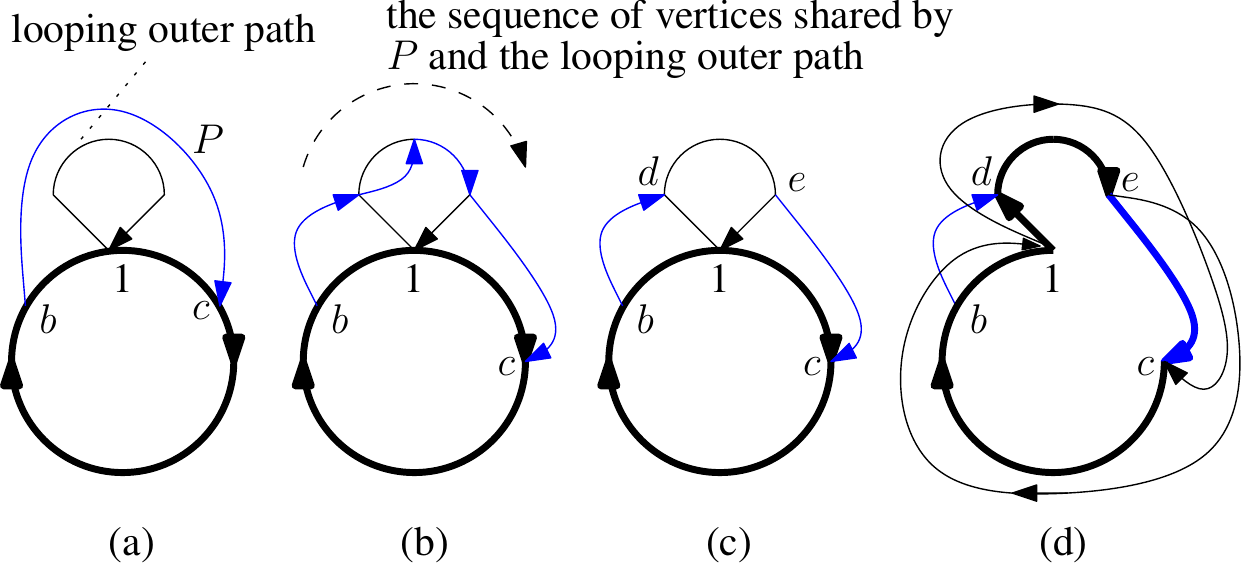}
\caption{Case 1 where there exists a looping outer path (drawn with thin black lines) that starts and ends at vertex~1. The center cycle $C_1$ is drawn with thick lines, and the second outer path (denoted as $P$) from $b$ to $c$, blue lines. To get another cycle after removing vertex~1, we must have that $1 < c \leq b \leq |V(C_1)|$, as shown in subfigure~(a). However, there are two vertex-disjoint cycles in subfigure~(a). So, $P$ must touch the looping outer path, as shown in subfigure (b). Taking the segment of $P$ from $C_1$ to the looping outer path, and that from the looping outer path back to $C_1$, we have subfigure (c). We can re-draw the path from $1$ to $c$ and that from $e$ to $1$ in subfigure (c) to get subfigure (d), where we have drawn the new center cycle with thick lines.}
\label{fig:looping}
\figgap
\end{figure}

\begin{IEEEproof}
{[}The \emph{only if} part:] We remove vertex~1. If there is another cycle, then there is a vertex (not vertex~1) in $C_1$ that has a path back to itself (this is because any cycle must share some vertex with $C_1$). This cannot happen if every outer path terminates at a higher-indexed vertex (we can ignore all outer paths that originate or terminate at vertex~1 as the vertex has been removed). So, there must exist an outer path with $b \geq c$.

[The \emph{if} part:] Clearly, if $b = c$, we have another cycle formed by the outer path. Otherwise, i.e., $ b > c$, the outer path and the path along $C_1$ from $c$ to $b$ form a cycle. See Figure~\ref{fig:looping}(a) for an example.
\end{IEEEproof}

Next, we define a \emph{looping} outer path as an outer path that originates and terminates at the same vertex in $C_1$. The graph $G_\text{sub}$ can be categorized as follows:
\begin{itemize}
\item there exists one or more looping outer path (Case 1), or
\item there is no looping outer path (which we will further divide into Cases 2 and 3).
\end{itemize}

We will show that in any case, we have Figure~\ref{fig:three-cycles}.

\subsection{Case 1: There Exists a Looping Outer Path}
Suppose that there exists a looping outer path from and to vertex~$1 \in V(C_1)$. This incurs no loss of generality as the choice of vertex~1 is arbitrary. Removing vertex~1 disconnects cycle $C_1$ and the cycle formed by the looping outer path. Recall that we need to remove two vertices to disconnect all cycles in $G_\text{sub}$. So, there must exist another cycle in $G_\text{sub}$. 

From Proposition~\ref{proposition:outer-path}, there exists another outer path $P$ from $b \in V(C_1) \setminus \{1\}$ to $c \in V(C_1) \setminus \{1\}$, where $b \geq c$. The outer path $P$ must share some vertex with the looping outer path; otherwise there exist two cycles as shown in Figure~\ref{fig:looping}(a).

Re-label the inner vertices of the looping outer path in ascending order, as follows: $1 \rightarrow (|V(C_1)|+1) \rightarrow (|V(C_1)|+2) \rightarrow \dotsm \rightarrow (|V(C_1)|+L) \rightarrow 1$, where $L$ is the number of inner vertices. It follows that the sequence of vertices shared by $P$ and the looping outer path (in the order of the direction of $P$) must be in ascending order (see Figure~\ref{fig:looping}(b)); otherwise, a cycle forms outside $C_1$.

See Figure~\ref{fig:looping}(c). Consider only the following segments of $P$: (i) from $b$ to the vertex where $P$ first touches the looping outer path, denoted by $d$; and (ii) the vertex where $P$ leaves the looping outer path, denoted by $e$, to $c$. It follows that $d \leq e$. By construction, all paths in Figure~\ref{fig:looping}(c) do not share inner vertices, i.e., they touch only at end points.  
Finally, re-draw Figure~\ref{fig:looping}(c) to get Figure~\ref{fig:looping}(d), which is isomorphic to Figure~\ref{fig:three-cycles}.

Note that vertices~1, $b$, and $d$ must be unique. 
We have shown that if there is a looping outer path, then we have the configuration in Figure~\ref{fig:three-cycles}, where path $I$ has zero arc, paths $W$ and $U$ possibly have zero arc (if $b=c$ and/or $d=e$), and all other paths must contain at least one arc.

\subsection{No Looping Outer Path}

For a non-looping outer path from vertex $b \in V(C_1)$ to $c \in V(C_1) \setminus \{b\}$, we say that the vertices \underline{in $C_1$} from $b$ to $c$ (in the direction of the arcs in $C_1$) but excluding $b$ and $c$ is \emph{covered} by this outer path. See Figure~\ref{fig:redraw}(b) for an example. 

For the purpose of this paper, we exclude outer paths with strictly smaller coverage, or multiple outer paths with equal coverage. Referring to Figure~\ref{fig:redraw}(c), consider an outer path that originates from $b$. Suppose that it has multiple paths back to $C_1$. We consider only the path (back to $C_1$) that has the \emph{largest coverage}. Similarly, for any path that terminates at $c$, we consider only the path (leaving $C_1$) that has the largest coverage. By doing this, each path that we consider has a unique originating vertex and a unique terminating vertex.

We now show the following property:
\begin{proposition} \label{proposition:full-coverage}
If there is no looping outer paths in $G_\text{sub}$, then all largest-covering outer paths must, together, provide full coverage for the cycle $C_1$. In other words, every vertex in $C_1$ must be covered by some outer path(s).
\end{proposition}

\begin{IEEEproof}
Consider any vertex $a \in V(C_1)$. Re-label $a$ as vertex~1, and other vertices $V(C_1)$ in ascending order in the arc direction. Remove vertex~1 from $G_\text{sub}$. There must exist another cycle.  It follows from Proposition~\ref{proposition:outer-path} that an outer path $P$ from $b$ to $c$ must exist, where $1 < c < b \leq |V(C_1)|$  ($c \neq b$ since there is no looping path), meaning that this outer path must cover vertex~1. We can safely ignore other outer paths that provide smaller or equal coverage, because if $P$ does not cover vertex~1, then none of the ignored outer paths does. Since the choice of $a$ is arbitrary, we have Proposition~\ref{proposition:full-coverage}.
\end{IEEEproof}

For example, the outer paths in Figure~\ref{fig:redraw}(d) provides full coverage for $C_1$, but the outer paths in Figures~\ref{fig:redraw}(b)--(c) do not. Removing one uncovered vertex from $C_1$ makes $G_\text{sub}$ acyclic.

Now, we consider $G_\text{sub}$ that consists of the cycle $C_1$ and all outer paths that provide the largest coverage (i.e., we remove all other arcs have gives smaller or equal coverage).  We are ready to proceed with Cases 2 and 3:
\begin{itemize}
\item There is no looping outer path, and no two outer paths have any common inner vertex (Case 2).
\item There is no looping outer path, and there exist two outer paths sharing the same inner vertex (Case 3).
\end{itemize}

\subsection{Case 2: No Looping Outer Path, and All Outer Paths Do Not Share Inner Vertices} 
We will show that we can always choose three outer paths to provide full coverage.

\begin{figure}[t]
\centering
\includegraphics[width=8.5cm]{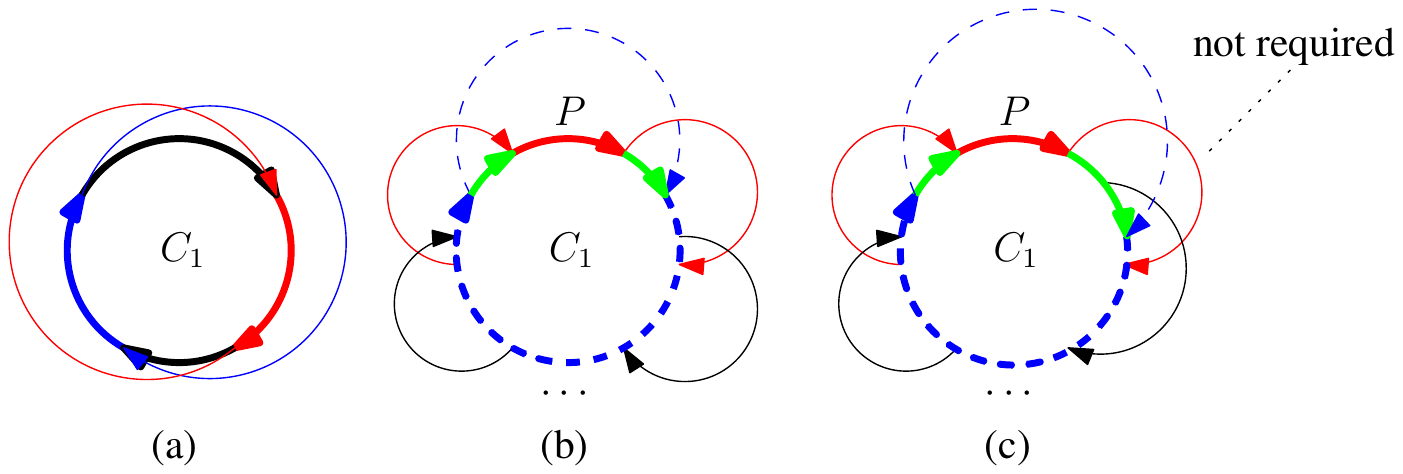}
\caption{(a) If two outer paths (drawn using thin lines) provide full coverage, we can always form two disjoint cycles, one marked with red and the other blue. (b) $G_\text{sub}$ with $K \geq 4$ outer paths providing full coverage can be converted to $K-2$ outer paths providing full coverage. (c) If the blue and the right black outer paths (non-adjacent) give overlapping coverage, then the right red outer path is actually redundant, i.e., $K-1$ outer path is sufficient to give full coverage, instead of $K$.}
\label{fig:2-coverage}
\figgap
\end{figure}

First, note that one outer path cannot provide full coverage.
Now, suppose that we can find two outer paths providing full coverage. We show in Figure~\ref{fig:2-coverage}(a) that we can always form two vertex-disjoint cycles. So, this also cannot happen.

Next, suppose that we can find three outer paths providing full coverage, we have Figure~\ref{fig:three-cycles}. As there is no looping outer path, the nine paths in Figure~\ref{fig:three-cycles} each have one or more arcs.


Finally, we show that if we can find $K \geq 4$ outer paths providing full coverage, we can always modify the cycles such that $(K-2)$ outer paths provide full coverage. We illustrate this in Figure~\ref{fig:2-coverage}(b). We do the following:
\begin{enumerate}
\item Combine the dashed blue arrows to be the new $C_1$.
\item Combine the  two adjacent (red) outer paths, and the red arc in $C_1$ that connects the two red outer paths (i.e., $P$, which can be of zero length) into a new outer path.
\item Remove the two green paths in $C_1$. Each green path must contain at least one arc; otherwise, the outer paths cannot provide full coverage.
\end{enumerate}
Note that by doing this, the new graph still retains the structure of a cycle with outer paths covering it. The new graph has $K-2$ outer paths providing full coverage. This reduction is always possible as the coverage of two non-adjacent outer paths does not overlap, illustrated in Figure~\ref{fig:2-coverage}(c).

By repeating this step, starting from any $K \geq 4$ outer paths, we can find a graph with $K=2$ or $K=3$ outer paths. As $K=2$ is not possible, we will always get a graph with $K=3$ outer paths providing full coverage, i.e., Figure~\ref{fig:three-cycles}.

\subsection{ Case 3: No Looping Outer Path and Two Outer Paths Share Some Inner Vertices} \label{sec:figure-end}
Let the two outer paths that share some common inner vertex be $P$ and $Q$, and one of the shared inner vertices be $z$.  Further, let the originating and terminating vertices of $P$ be $p_1$ and $p_\text{last}$ respectively, and those of $Q$ be $q_1$ and $q_\text{last}$. Here, $p_1 \neq p_\text{last}$ and $q_1 \neq q_\text{last}$ as there is no looping outer path, and $p_1 \neq q_1$ and $p_\text{last} \neq q_\text{last}$ as no two outer paths have the same originating or terminating vertices.

\begin{figure}[t]
\centering
\includegraphics[width=8cm]{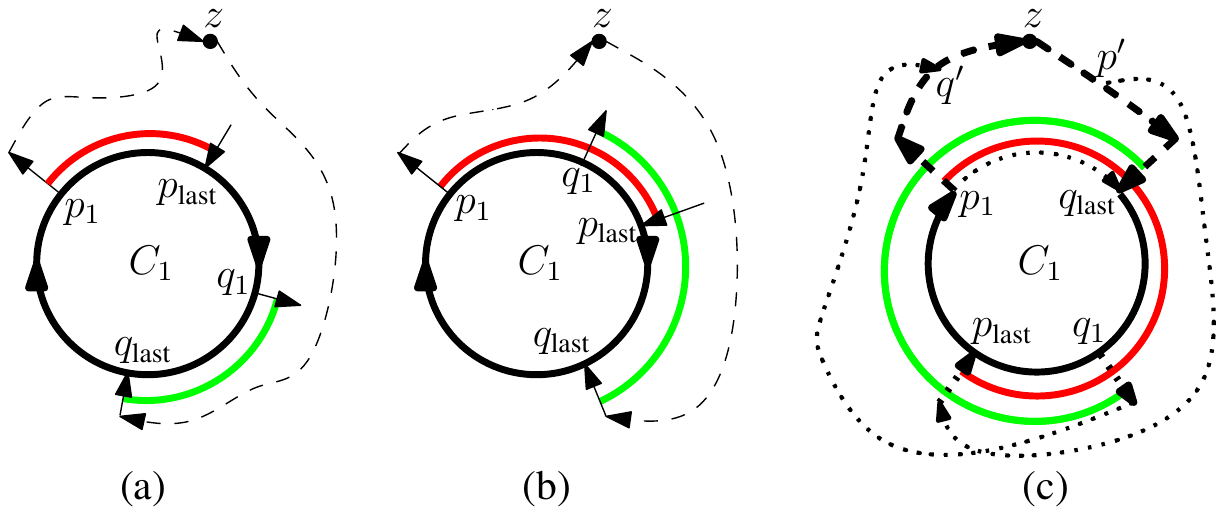}
\caption{The overlapping of the coverage of two outer paths, where the red line represents the coverage of the outer path $P$ ($p_1 \rightarrow \dotsm \rightarrow p_\text{last}$), and the green line the outer path $Q$ ($q_1 \rightarrow \dotsm \rightarrow q_\text{last}$)}
\label{fig:overlapping-coverage}
\figgap
\end{figure}

Now, the coverage of $P$ and $Q$ can be either (a) non-overlapping, (b) overlapping once, or (c) overlapping twice, as shown in Figure~\ref{fig:overlapping-coverage}. The red line shows the coverage of $P$, and the green line $Q$. By definition, there is a subpath from $p_1$ to $z$ along $P$ and another subpath from $z$ to $p_\text{last}$ along $P$. The two subpaths must be vertex-disjoint, except $z$, as there is no cycle in $P$.  Similarly, we have two vertex-disjoint paths from $q_1$ to $z$, and from $z$ to $q_\text{last}$, both along $Q$. This means, there is an subpath from $p_1$ to $q_\text{last}$ through $z$, and another from $q_1$ to $p_\text{last}$ through $z$. So, $p_1 \neq q_\text{last}$, $q_1 \neq p_\text{last}$, as there is no looping outer path, and hence $p_1$, $p_\text{last}$, $q_1$, and $q_\text{last}$ are distinct. 

Suppose that we have Figure~\ref{fig:overlapping-coverage}(a). The largest-covering outer path from $p_1$ should terminate at $q_\text{last}$, and that from $q_1$ at $p_\text{last}$. The outer path from $p_1$ to $q_\text{last}$ and that from $q_1$ to $p_\text{last}$ should have been chosen. This means the largest-covering paths actually overlap twice, i.e., we should have Figure~\ref{fig:overlapping-coverage}(c).

Suppose that we have Figure~\ref{fig:overlapping-coverage}(b). The outer path from $p_1$ to $q_\text{last}$, through $z$, gives the largest coverage, and it would have been chosen. 

So, we can only have the configuration in Figure~\ref{fig:overlapping-coverage}(c), where the coverage overlaps twice. The coverage from $p_1$ to $q_\text{last}$ is smaller than that from $p_1$ to $p_\text{last}$. So, the largest-covering outer path from $p_1$ was correctly identified. Similarly, the largest-covering outer path from $q_1$ terminates at $q_\text{last}$.


We will now show that we can always get Figure~\ref{fig:three-cycles} from Figure~\ref{fig:overlapping-coverage}(c). Recall that there is a subpath from $p_1$ to $z$ and another subpath from $z$ to $q_\text{last}$, and these two subpaths are vertex-disjoint, except $z$. We denote the outer path from $p_1$ to $q_\text{last}$ (through $z$) by $Z$ (drawn with a thick dashed line). 

Next, recall that there is a subpath from $q_1$ to $z$, and another from $z$ to $p_\text{last}$. So, the subpath from $q_1$ to $z$ must meet $Z$. Denote the vertex it first meets $Z$ as $q'$. Similarly, the subpath from $z$ to $p_\text{last}$ must share some common vertices with $Z$ (at least vertex~$z$). Let the last shared vertex be $p'$. With this construction, $Z$, the subpath from $q_1$ to $q'$, and the subpath from $p'$ to $p_\text{last}$ are vertex-disjoint, except at  $p'$ and $q'$.

We now re-draw Figure~\ref{fig:overlapping-coverage}(c) as follows: Let the path from $q_\text{last}$ to $p_1$ along $C_1$ (drawn with a thick solid line) plus path $Z$ (drawn with a thick dashed line) be the center cycle, and let the subpaths (drawn with dotted arrows) (i) from $p_1$ to $q_\text{last}$ along $C_1$, (ii) from $p'$ to $p_\text{last}$, and (iii) from $q_1$ to $q'$ be the three outer paths. Note that only $p'$ and $q'$ can co-locate. This is isomorphic to Figure~\ref{fig:three-cycles}, with path $I$ possibly having zero arc (if $p' = q' = z$).

Combining the Cases 1--3, we have Lemma~\ref{lemma:special-subgraph}. $\hfill\blacksquare$

\section{Conclusion}

We have solved a new class of index-coding problems, characterized by their side-information graphs. 
We have shown that for any side-information graph whose maximum acyclic induced subgraph (MAIS) can be formed by removing two or fewer vertices, the optimal index codelength (i) equals the order of the MAIS, and (ii) is achievable by linear index codes.
We proved this by constructing a special subgraph that the side-information graph must contain, and design a linear index code on it. 
We then show that the liner index code achieves the MAIS lower bound.

We have incidentally characterized a subset of directed graphs whose minrank equals $|V(G)|-1$ or $|V(G)|-2$, where $|V(G)|$ is the order of the graph $G$.

The result of this paper has led to another recent result: for any side-information graph of up to five vertices, the optimal index codelength is achievable using linear index codes~\cite{ong14isitsubmitted}.

\end{document}